\documentclass[structabstract]{aa}
\usepackage{graphicx}
\usepackage{natbib}

\begin{document}

    \title{Kepler-9 revisited}
         \subtitle{$60\%$ the mass with six times more data}
    \author{Stefan Dreizler \and \inst{1} Aviv Ofir \inst{1},\inst{2} }

    \institute{Institut f\"ur Astrophysik, Georg-August-Universit\"at, 
Friedrich-Hund-Platz 1, 37077 G\"ottingen, Germany 
\and
Kepler Participating Scientist\\
\email{dreizler@astro.physik.uni-geottingen.de}}

    \date{Received XXX; accepted YYY}
   \abstract
{}
{Kepler-9 was the first case where transit timing variations
have been used to confirm the planets in this system. Following
predictions of dramatic TTVs - larger than a week - we re-analyse the
system based on the full Kepler data set.}  
{We re-processed
all available data for Kepler-9 removing short and long term trends,
measured the times of mid-transit and used those for dynamical
analysis of the system.}  
{The newly determined masses and
radii of Kepler-9b and -9c change the nature of these planets relative
to the one described in Holman et al. 2010 (hereafter H10) with very
low, but relatively well charcterised (to better than 7\%), bulk
densities of 0.18 and 0.14 g cm$^3$ (about 1/3 of the H10 value). We
constrain the masses (45.1 and 31.0 M$_\oplus$, for Kepler-9b and -9c
respectively) from photometry alone, allowing us to see possible
indications for an outer non-transiting planet in the radial velocity
data. At $2R_\oplus$ Kepler-9d is determined to be larger than
suggested before - suggesting that it is a low-mass low-density
planet.}  
{The comparison between the H10 analysis and our
new analysis suggests that small formal error in the TTV inversion may
be misleading if the data does not cover a significant fraction of the
interaction time scale.}


\titlerunning{Kepler-9 revisited}

\maketitle

\section{Introduction}
Transit timing variations (TTVs) are deviations from strict
periodicity in extra solar planetary transits, caused by non-Keplerian
forces -- usually the interaction with other planets in the
system. These TTVs are particularly important in multi-transiting
systems since they can allow learning about dynamics in the system,
which in turn can confirm the exoplanetary origin of the transit
signals with no further observations (e.g. \citealt{2010Sci...330...51H}, H10 hereafter, 
or \citealt{2013ApJS..208...22X}), and sometimes even allow deriving the planets' mass from 
photometry alone
\citep[Kepler-87,][]{2014A&A...561A.103O}. For these reasons TTVs had
attracted a lot of attention since they were first predicted by
\citet{2005Sci...307.1288H} and \citet{2005MNRAS.359..567A}, and
especially since they where first observed by H10 in the prototypical
Kepler-9 system.

Kepler-9 is prototypical not just because it was the first object
detected with TTVs, but also since it is a textbook-like example of
TTVs: exhibiting very large TTVs on very deep transits, making the
effect abundantly clear. The first study of the Kepler-9 system also
included a prediction for the expected TTVs during the following few
years (their Figure S4) which included dramatic TTV spanning up to
about $^{+4}_{-8}d$ relative to the nominal ephemeris, accumulated
over long interaction times scales (e.g. $\sim 1000d$ from first
maximum to first minimum TTV excursion of Kepler-9c). These very large
TTVs are easy to compare to the observed ones in later Kepler
data. Indeed by the time we re-analysed this object much more data were
available, revealing that the actual TTVs, while still large, were
much less extreme than initially predicted. We observed TTV spans of
about $^{+0.6}_{-0.9}d$ for the same features as above, and TTV time
scale about half as long as predicted. These prompted us to revisit
the analysis of Kepler-9. 

This paper is therfore divided in the following way: in sections \S
\ref{photometry} and \S\ref{modeling} we describe the input data and
TTV analysis procedures we used. In \S \ref{recovery} we made sure we
are able to recover the H10 results when using only the data that was
available at the time, showing consistent analysis, which then allowed
us to perform a full analysis using the full data set in \S
\ref{FullTTV}, before discussing the updated analysis in  \S
\ref{Discussion}.

\section{Photometry and light curve modeling}
\label{photometry}
We processed quarters 1 through 16 of Kepler-9 long cadence photometry 
which spans 1426d, more than six times longer than the original study 
of H10. We processed it similarly to the processing of Kepler-87
\citep{2014A&A...561A.103O}, fitting for Kepler-9b and Kepler-9c's
individual times of mid-transit, and fitting Kepler-9d using linear
ephemeris since we detected no TTVs for it. In short, this processing
used a light curve that was corrected for short-term systematic
effects using the SARS algorithm
\citep{2010MNRAS.404L..99O,2013A&A...555A..58O}, which was then
iteratively: fitted for long-term trends, modeled for its transit
signals, and corrected for the model by dividing it out -- till
convergence. The resultant photometric model parameters are given in Table
\ref{PhotometryTable}. 

\begin{table*}
\caption{All fitted parameters from the light curve modeling of the Kepler-9 
  system. Note a few
  non-fitted parameters are also given for convenience: the linear
  periods of Kepler-9b and Kepler-9c, which exhibit TTVs, are computed
  from the individual times of mid-transit, and the different
  semi-major axes $a$ are actually a single parameter common to all
  planets scaled using Kepler's laws. All times (T$_{mid}$ parameters)
  are measured relative to BJD-2454933.0.}
\label{PhotometryTable}
\begin{tabular}{|l l l l |l l l l|}
\hline
param       &     BestFit &   $50\%$   &    $1 \sigma$  range 		& param       &     BestFit &   $50\%$   &    $1 \sigma$  range \\
\hline 
$\mathrm{Linear}\,P_b[d]$ & 19.2471658 & 19.2471669 & $_{-0.0000036}^{+ 0.0000035}$& $Tmid_{b,47} $ & 987.78055 & 987.78027 & $_{  -0.00053}^{+   0.00050}$\\
$a_b/R_*   $ &      25.00 &      24.95 & $_{     -0.16}^{+      0.17}$& $Tmid_{b,48} $ & 1007.00387 & 1007.00382 & $_{  -0.00059}^{+   0.00055}$\\
$r_b/R_*   $ &   0.082483 &   0.082515 & $_{ -0.000107}^{+  0.000099}$& $Tmid_{b,49} $ & 1045.44241 & 1045.44326 & $\pm    0.00059$\\
$b_b/R_*   $ &     0.7803 &     0.7812 & $_{   -0.0036}^{+    0.0033}$& $Tmid_{b,50} $ & 1083.88091 & 1083.88077 & $_{  -0.00056}^{+   0.00059}$\\
$\mathrm{Linear}\,P_c[d]$ &  38.944030 &  38.944011 & $\pm   0.000012$& $Tmid_{b,51} $ & 1103.09894 & 1103.09864 & $\pm    0.00072$\\
$a_c/R_*   $ &      40.00 &      39.91 & $_{     -0.25}^{+      0.27}$& $Tmid_{b,52} $ & 1122.31856 & 1122.31921 & $_{  -0.00062}^{+   0.00057}$\\
$r_c/R_*   $ &    0.07963 &    0.07964 & $_{  -0.00015}^{+   0.00013}$& $Tmid_{b,53} $ & 1141.53855 & 1141.53895 & $\pm    0.00061$\\
$b_c/R_*   $ &     0.8619 &     0.8624 & $_{   -0.0023}^{+    0.0019}$& $Tmid_{b,54} $ & 1160.75793 & 1160.75827 & $_{  -0.00062}^{+   0.00058}$\\
$P_d       $ & 1.59295922 & 1.59295878 & $_{-0.00000095}^{+0.00000109}$& $Tmid_{b,55} $ & 1179.97760 & 1179.97817 & $_{  -0.00055}^{+   0.00058}$\\
$\mathrm{Linear}\,T_{mid,3}$ &  800.15785 &  800.15795 & $\pm    0.00026$& $Tmid_{b,56} $ & 1199.19677 & 1199.19654 & $_{  -0.00056}^{+   0.00054}$\\
$a_d/R_*   $ &      4.748 &      4.738 & $_{    -0.030}^{+     0.032}$& $Tmid_{b,57} $ & 1218.41779 & 1218.41789 & $_{  -0.00064}^{+   0.00062}$\\
$r_d/R_*   $ &    0.01508 &    0.01517 & $\pm    0.00013$	      & $Tmid_{b,58} $ & 1237.63676 & 1237.63612 & $_{  -0.00045}^{+   0.00043}$\\
$b_d/R_*   $ &     0.6955 &     0.6951 & $_{   -0.0075}^{+    0.0064}$& $Tmid_{b,59} $ & 1256.86020 & 1256.86036 & $_{  -0.00054}^{+   0.00057}$\\
\cline{1-4} $Tmid_{b,1}  $ &   44.24962 &   44.24968 & $_{  -0.00054}^{+   0.00053}$& $Tmid_{b,60} $ & 1276.08020 & 1276.07952 & $\pm    0.00053$\\
$Tmid_{b,2}  $ &   63.48423 &   63.48436 & $_{  -0.00069}^{+   0.00066}$& $Tmid_{b,61} $ & 1395.30246 & 1295.30274 & $_{  -0.00071}^{+   0.00070}$\\
$Tmid_{b,3}  $ &  101.95597 &  101.95545 & $_{  -0.00059}^{+   0.00053}$& $Tmid_{b,62} $ & 1333.74328 & 1333.74329 & $_{  -0.00051}^{+   0.00052}$\\
$Tmid_{b,4}  $ &  121.19173 &  121.19124 & $_{  -0.00055}^{+   0.00067}$& $Tmid_{b,63} $ & 1352.96620 & 1352.96606 & $\pm    0.00051$\\
$Tmid_{b,5}  $ &  140.43520 &  140.43504 & $_{  -0.00089}^{+   0.00106}$& $Tmid_{b,64} $ & 1372.19038 & 1372.18961 & $_{  -0.00091}^{+   0.00089}$\\
$Tmid_{b,6}  $ &  178.92599 &  178.92624 & $_{  -0.00051}^{+   0.00049}$& $Tmid_{b,65} $ & 1391.41097 & 1391.41048 & $_{  -0.00053}^{+   0.00060}$\\
$Tmid_{b,7}  $ &  198.17235 &  198.17249 & $_{  -0.00053}^{+   0.00046}$& $Tmid_{b,66} $ & 1410.63395 & 1410.63366 & $_{  -0.00053}^{+   0.00051}$\\
$Tmid_{b,8}  $ &  217.42998 &  217.42953 & $_{  -0.00055}^{+   0.00053}$& $Tmid_{b,67} $ & 1429.85806 & 1429.85854 & $_{  -0.00053}^{+   0.00054}$\\
$Tmid_{b,9}  $ &  236.68239 &  236.68220 & $_{  -0.00056}^{+   0.00060}$& $Tmid_{b,68} $ & 1449.08463 & 1449.08471 & $_{  -0.00069}^{+   0.00067}$\\
\cline{5-8} $Tmid_{b,10} $ &  255.94584 &  255.94603 & $_{  -0.00055}^{+   0.00062}$& $Tmid_{c,1}  $ &   36.30566 &   36.30599 & $_{  -0.00076}^{+   0.00078}$\\
$Tmid_{b,11} $ &  275.20392 &  275.20414 & $_{  -0.00054}^{+   0.00051}$& $Tmid_{c,2}  $ &   75.33116 &   75.33166 & $_{  -0.00069}^{+   0.00073}$\\
$Tmid_{b,12} $ &  294.47506 &  294.47509 & $_{  -0.00059}^{+   0.00057}$& $Tmid_{c,3}  $ &  114.33665 &  114.33623 & $_{  -0.00086}^{+   0.00084}$\\
$Tmid_{b,13} $ &  313.73752 &  313.73698 & $_{  -0.00066}^{+   0.00051}$& $Tmid_{c,4}  $ &   153.3191 &   153.3201 & $_{   -0.0011}^{+    0.0012}$\\
$Tmid_{b,14} $ &  333.01416 &  333.01480 & $_{  -0.00060}^{+   0.00062}$& $Tmid_{c,5}  $ &  192.26400 &  192.26409 & $_{  -0.00103}^{+   0.00094}$\\
$Tmid_{b,15} $ &  352.28289 &  352.28264 & $_{  -0.00056}^{+   0.00053}$& $Tmid_{c,6}  $ &  231.18287 &  231.18412 & $_{  -0.00087}^{+   0.00082}$\\
$Tmid_{b,16} $ &  371.56419 &  371.56349 & $_{  -0.00054}^{+   0.00055}$& $Tmid_{c,7}  $ &  270.07285 &  270.07225 & $_{  -0.00076}^{+   0.00071}$\\
$Tmid_{b,17} $ &  390.83565 &  390.83622 & $_{  -0.00063}^{+   0.00059}$& $Tmid_{c,8}  $ &  308.92941 &  308.93019 & $_{  -0.00075}^{+   0.00073}$\\
$Tmid_{b,18} $ &  410.11896 &  410.11837 & $_{  -0.00058}^{+   0.00060}$& $Tmid_{c,9}  $ &  347.76595 &  347.76565 & $_{  -0.00072}^{+   0.00070}$\\
$Tmid_{b,19} $ &  429.39382 &  429.39413 & $_{  -0.00069}^{+   0.00074}$& $Tmid_{c,10} $ &  386.57832 &  386.57828 & $_{  -0.00071}^{+   0.00069}$\\
$Tmid_{b,20} $ &  448.67952 &  448.68016 & $_{  -0.00060}^{+   0.00061}$& $Tmid_{c,11} $ &  425.37752 &  425.37789 & $_{  -0.00073}^{+   0.00070}$\\
$Tmid_{b,21} $ &  467.95695 &  467.95682 & $_{  -0.00055}^{+   0.00057}$& $Tmid_{c,12} $ &  464.16924 &  464.16864 & $_{  -0.00087}^{+   0.00090}$\\
$Tmid_{b,22} $ &  487.24089 &  487.24071 & $_{  -0.00057}^{+   0.00059}$& $Tmid_{c,13} $ &  502.95866 &  502.95890 & $_{  -0.00076}^{+   0.00075}$\\
$Tmid_{b,23} $ &   506.5199 &   506.5196 & $_{   -0.0013}^{+    0.0016}$& $Tmid_{c,14} $ &  541.75376 &  541.75425 & $_{  -0.00105}^{+   0.00099}$\\
$Tmid_{b,24} $ &  525.80356 &  525.80302 & $_{  -0.00087}^{+   0.00081}$& $Tmid_{c,15} $ &  580.56155 &  580.56190 & $\pm    0.00072$\\
$Tmid_{b,25} $ &  545.07889 &  545.07896 & $_{  -0.00054}^{+   0.00062}$& $Tmid_{c,16} $ &  619.38606 &  619.38610 & $_{  -0.00086}^{+   0.00100}$\\
$Tmid_{b,26} $ &  564.36033 &  564.36037 & $_{  -0.00059}^{+   0.00062}$& $Tmid_{c,17} $ &  658.23669 &  658.23613 & $_{  -0.00084}^{+   0.00082}$\\
$Tmid_{b,27} $ &  583.63569 &  583.63578 & $_{  -0.00056}^{+   0.00059}$& $Tmid_{c,18} $ &  697.11331 &  697.11376 & $_{  -0.00084}^{+   0.00081}$\\
$Tmid_{b,28} $ &  602.90982 &  602.90991 & $_{  -0.00051}^{+   0.00054}$& $Tmid_{c,19} $ &  736.02270 &  736.02225 & $_{  -0.00082}^{+   0.00079}$\\
$Tmid_{b,29} $ &  641.45176 &  641.45165 & $_{  -0.00059}^{+   0.00064}$& $Tmid_{c,20} $ &  713.93519 &  813.93487 & $_{  -0.00073}^{+   0.00075}$\\
$Tmid_{b,30} $ &   660.7231 &   660.7215 & $_{   -0.0024}^{+    0.0015}$& $Tmid_{c,21} $ &  752.93504 &  852.93585 & $_{  -0.00069}^{+   0.00075}$\\
$Tmid_{b,31} $ &  679.98214 &  679.98182 & $_{  -0.00076}^{+   0.00078}$& $Tmid_{c,22} $ &  791.96021 &  891.95978 & $_{  -0.00085}^{+   0.00084}$\\
$Tmid_{b,32} $ &  699.24308 &  699.24380 & $_{  -0.00054}^{+   0.00057}$& $Tmid_{c,23} $ & 931.00067 & 931.00056 & $_{  -0.00071}^{+   0.00064}$\\
$Tmid_{b,33} $ &  718.49797 &  718.49867 & $_{  -0.00057}^{+   0.00061}$& $Tmid_{c,24} $ & 970.05421 & 970.05463 & $_{  -0.00095}^{+   0.00089}$\\
$Tmid_{b,34} $ &  737.75264 &  737.75284 & $_{  -0.00070}^{+   0.00069}$& $Tmid_{c,25} $ & 1009.11603 & 1009.11728 & $\pm    0.00079$\\
$Tmid_{b,35} $ &  757.00132 &  757.00190 & $_{  -0.00062}^{+   0.00059}$& $Tmid_{c,26} $ & 1048.18462 & 1048.18524 & $\pm    0.00079$\\
$Tmid_{b,36} $ &  776.24954 &  776.24997 & $_{  -0.00059}^{+   0.00053}$& $Tmid_{c,27} $ & 1087.25806 & 1087.25684 & $_{  -0.00074}^{+   0.00079}$\\
$Tmid_{b,37} $ &  795.49196 &  795.49128 & $_{  -0.00053}^{+   0.00054}$& $Tmid_{c,28} $ & 1126.33190 & 1126.33108 & $_{  -0.00073}^{+   0.00075}$\\
$Tmid_{b,38} $ &  814.73192 &  814.73155 & $_{  -0.00065}^{+   0.00076}$& $Tmid_{c,29} $ & 1165.40489 & 1165.40450 & $_{  -0.00076}^{+   0.00071}$\\
$Tmid_{b,39} $ &  833.96754 &  833.96814 & $_{  -0.00066}^{+   0.00064}$& $Tmid_{c,30} $ & 1204.47870 & 1204.47888 & $_{  -0.00099}^{+   0.00105}$\\
$Tmid_{b,40} $ &  853.20484 &  853.20451 & $_{  -0.00059}^{+   0.00057}$& $Tmid_{c,31} $ & 1243.55397 & 1243.55348 & $_{  -0.00081}^{+   0.00071}$\\
$Tmid_{b,41} $ &  872.43257 &  872.43290 & $_{  -0.00054}^{+   0.00057}$& $Tmid_{c,32} $ & 1282.62718 & 1282.62584 & $_{  -0.00066}^{+   0.00067}$\\
$Tmid_{b,42} $ &  891.66331 &  891.66329 & $\pm    0.00057$& $Tmid_{c,33} $ & 1321.69645 & 1321.69632 & $_{  -0.00081}^{+   0.00076}$\\
$Tmid_{b,43} $ &  910.88965 &  910.88962 & $_{  -0.00050}^{+   0.00048}$& $Tmid_{c,34} $ & 1360.76777 & 1360.76784 & $_{  -0.00067}^{+   0.00069}$\\
$Tmid_{b,44} $ &  930.11432 &  930.11491 & $_{  -0.00064}^{+   0.00061}$& $Tmid_{c,35} $ & 1399.83472 & 1399.83481 & $_{  -0.00076}^{+   0.00073}$\\
$Tmid_{b,45} $ &  949.33779 &  949.33794 & $_{  -0.00051}^{+   0.00050}$& $Tmid_{c,36} $ & 1438.90270 & 1438.90226 & $_{  -0.00074}^{+   0.00071}$\\
$Tmid_{b,46} $ &  968.56094 &  968.56129 & $_{  -0.00054}^{+   0.00056}$& 	     & 		  & 	       & 			      \\
\hline
\end{tabular}
\end{table*}

\section{TTV modeling}
\label{modeling}
We did not include Kepler-9d in the TTV modeling since it is dynamically 
decoupled from the outer two planets (see also Section\,\ref{FullTTV}). 
For the modeling
of the TTVs we use the {\it mercury6} code
\citep{1997DPS....29.2706C,1999MNRAS.304..793C}. The integration of
the planetary orbits is done using the Bulirsch-Stoer integrator
implemented in {\it mercury6} 
starting from a set of initial values for the orbital elements for the
planetary system. For integration of orbits in the Kepler-9 system,
we use a time step of 0.5\,days, i.e. about one 40$^{\mathrm th}$ of
the orbital period of planet b. The duration of the integration is
limited to the duration of the Kepler mission. From the osculating
orbital elements at each time step we calculate the next transit
time. The final transit times are then calculated from spline
interpolations. These calculated and the observed mid transit times
are used to run a Levenberg-Marquardt optimization resulting in an
optimized parameter set for the planetary system.

Since the fit may depend on the choice of the initial values, we use
the best fit parameters as well as the formal fit errors from the
covariance matrix for a second extended fit. Within the 3-$\sigma$
limit we randomly vary the start parameters, however obeying parameter
limits, e.g. positive eccentricity, if required. This procedure probes
the $\chi^2$-landscape around the initial best fit value, it typically
finds a better best fit and we use the distribution of parameters as
an estimate of the error bars.

As a final check, we also integrate the best fit orbital solution
over 5\,Gyr using the hybrid-symplectic integrator of {\it mercury6}
at a time step of 0.8 days. Only a long-term stable solution is
accepted.

\section{Recovery of previous results}
\label{recovery}
In a first step, we use the TTV data from H10 in order to demonstrate
that we can recover their solution. Given the rather low number of TTV
measurements, we restrict the orbits to coplanar orbits, given the low
dispersion in measured inclinations that seems not to be a restrictive
constraint. The free parameters therefore are the orbital period,
eccentricity, argument of periastron, and mean anomaly at the
beginning of the integration for each of the two planets. We take the
mass of the central star as input with a distribution according to
\citet{2011A&A...531A...3H}. During each fit, the stellar mass is
fixed. Instead of using the planetary masses as parameters, the mass
of planet b is given relative to the stellar mass, the mass of planet
c relative to that of planet b. We use 2500 random starting values for
the Levenberg-Marquardt optimization as described in
Sect.\,\ref{modeling}. As also discussed by H10, the planetary masses
can only be weakly constrained from the partial TTV data set. We
therefore also included the radial velocity (RV) measurements from H10
in our fit for the partial data set. 

\begin{table}
\caption{Parameters of the planetary system of Kepler-9 derived from
  the TTV analysis using the partial data set of H10. The stellar mass
  is an input parameter with a distribution according
  \citet{2011A&A...531A...3H}.The osculating orbital 
  elements are given at a reference time BJD=2454900.0. For
  comparison, we list the literature value for the stellar mass 
  \citep{2011A&A...531A...3H} and those of the the planetary parameters 
  from H10.}
\label{ResultsOld}
\begin{tabular}{l l l l l}
\hline
\hline
Parameter              & \multicolumn{2}{c}{this work} & \multicolumn{2}{c}{Holman et al.} \\
                       & best fit      & $\sigma$ &  best fit & $\sigma$ \\    
\hline
m$_{\rm b}$ [M$_\oplus$]    &  79.6$^{*}$      &  3.6      & 79.9$^{*}$  & 6.5\\
m$_{\rm c}$ [M$_\oplus$]    &  54.8$^{*}$      &  2.6      & 54.4       & 4.1 \\
m$_{\rm c}$/m$_{\rm b}$     & 0.688           & 0.004       & 0.680      & 0.02 \\
P$_{\rm b}$ [days]         & 19.2159         &   0.0008  & 19.2372    & 0.0007\\
P$_{\rm c}$ [days]         & 39.084          &   0.003   & 38.992     & 0.005 \\
a$_{\rm b}$ [AU]           & 0.143$^{*}$     &  0.001     & 0.140$^{*}$ & 0.001\\
a$_{\rm c}$ [AU]           & 0.229$^{*}$     &   0.002    & 0.225$^{*}$ & 0.001 \\
e$_{\rm b}$                & 0.10           &   0.02     & 0.15$^{*}$   & 0.04\\
e$_{\rm c}$                & 0.08           &   0.02     & 0.13$^{*}$   & 0.04\\
$\omega_{\rm b}$ [$^\circ$] & 357.5          &  21.0      & 18.6$^{*}$   & 1.2\\
$\omega_{\rm c}$ [$^\circ$] & 101.5          &   4.0      & 101.3$^{*}$  & 9.6\\
\hline
                        & \multicolumn{2}{c}{this work} & \multicolumn{2}{c}{Havel et al.} \\
m$_\star$ [M$_\odot$]     & 1.05$^{+}$    & 0.03  & 1.05 & 0.03 \\
\hline
\multicolumn{5}{l}{$^{+}$ input value} \\
\multicolumn{5}{l}{$^{*}$ derived, i.e. not fitted parameters} \\
\multicolumn{5}{l}{note that H10 use $e\cos\omega$ and $e\sin\omega$} \\
\end{tabular}
\end{table}

\begin{figure}
\includegraphics[width=0.5\textwidth]{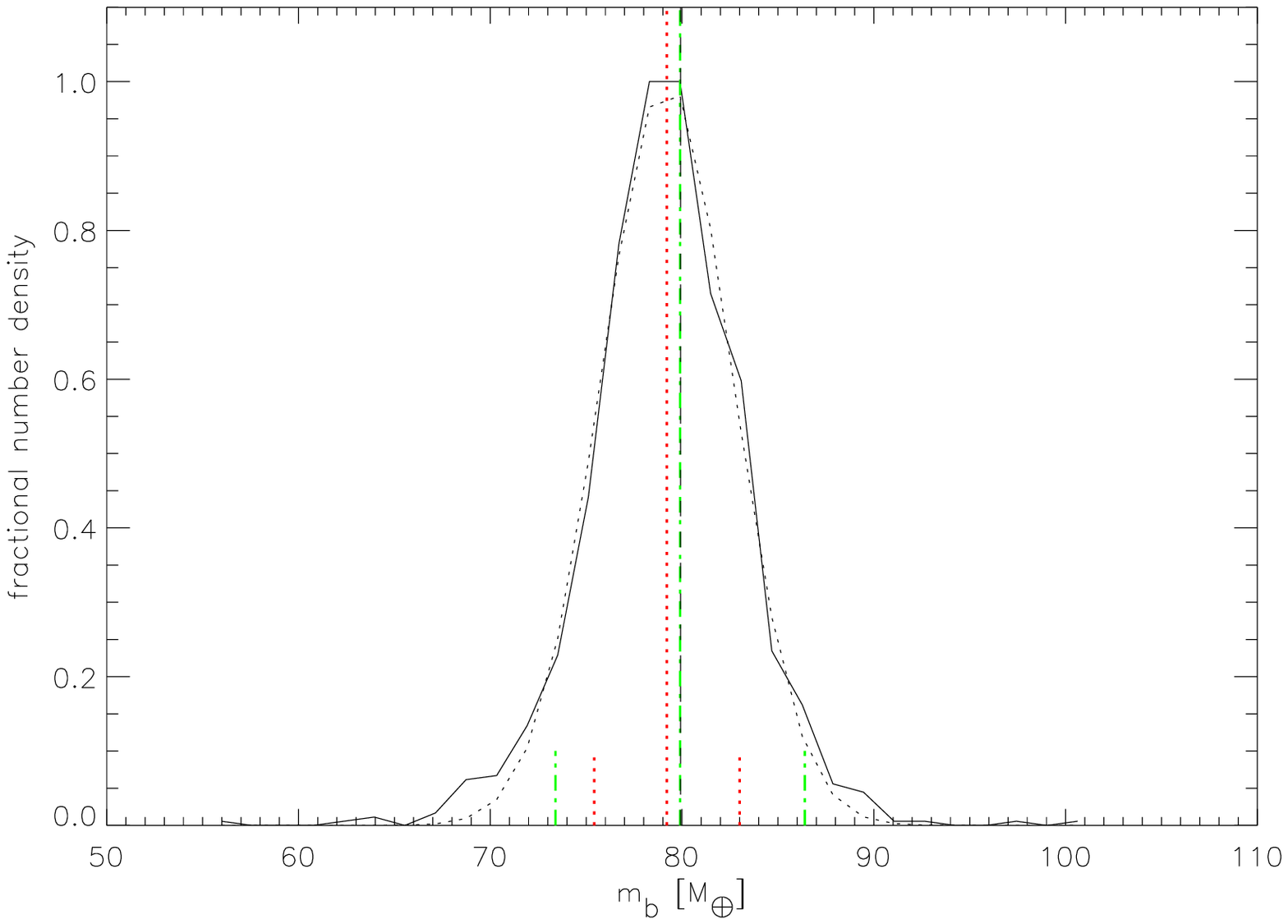}
\caption{Distribution of derived mass of planet b (black
  solid line)derived using the partial
  data set. The Gauss fit to this distribution is shown as black
  dotted line. The best fit as well as the $1\sigma$ error range is
  indicated as red dotted line. The median of the distribution is
  indicated as black dashed line. The planetary mass and its
  uncertainty derived by H10 is indicated as green dashed-dotted line.}
\label{FigMassP1Old}
\end{figure}

\begin{figure}
\includegraphics[width=0.5\textwidth]{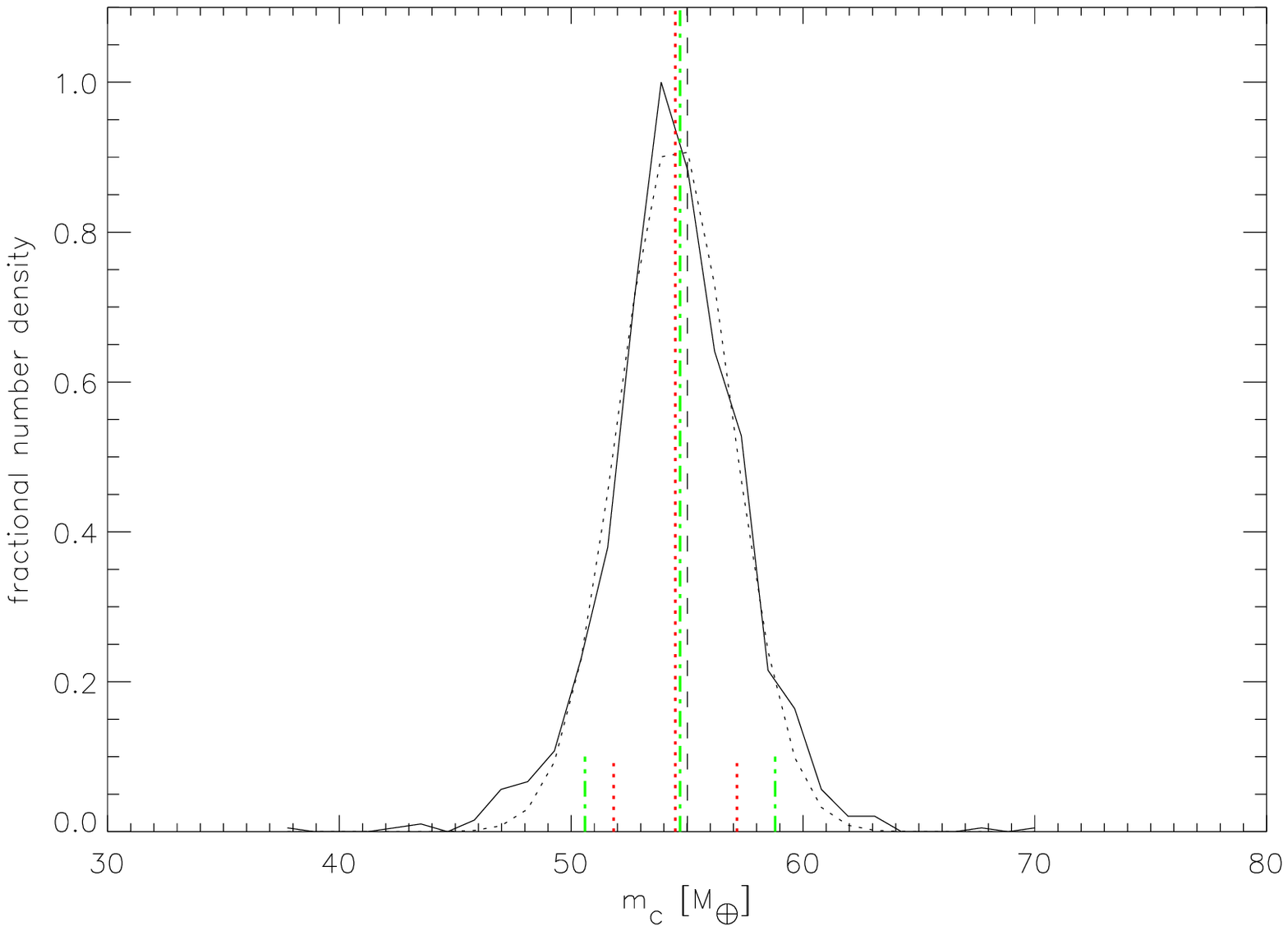}
\caption{Distribution of derived mass of planet c (black
  solid line) derived using the partial
  data set. The Gauss fit to this distribution is shown as black
  dotted line. The best fit as well as the $1\sigma$ error range is
  indicated as red dotted line. The median of the distribution is
  indicated as black dashed line. The planetary mass and its
  uncertainty derived by H10 is indicated as green dashed-dotted line.}
\label{FigMassP2Old}
\end{figure}

\begin{figure}
\includegraphics[width=0.5\textwidth]{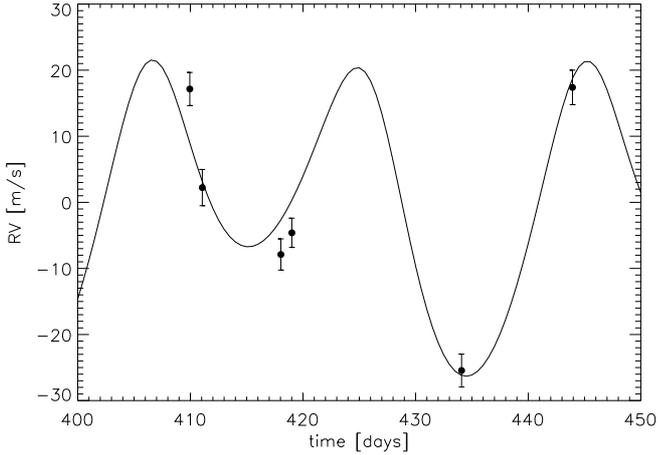}
\caption{Radial velocity measurements from H10 compared to those
  predicted by our best fit for the partial data set. The time is given as 
  BJD - 2454933.0.}  
\label{FigRV}
\end{figure}

\begin{figure}
\includegraphics[width=0.5\textwidth]{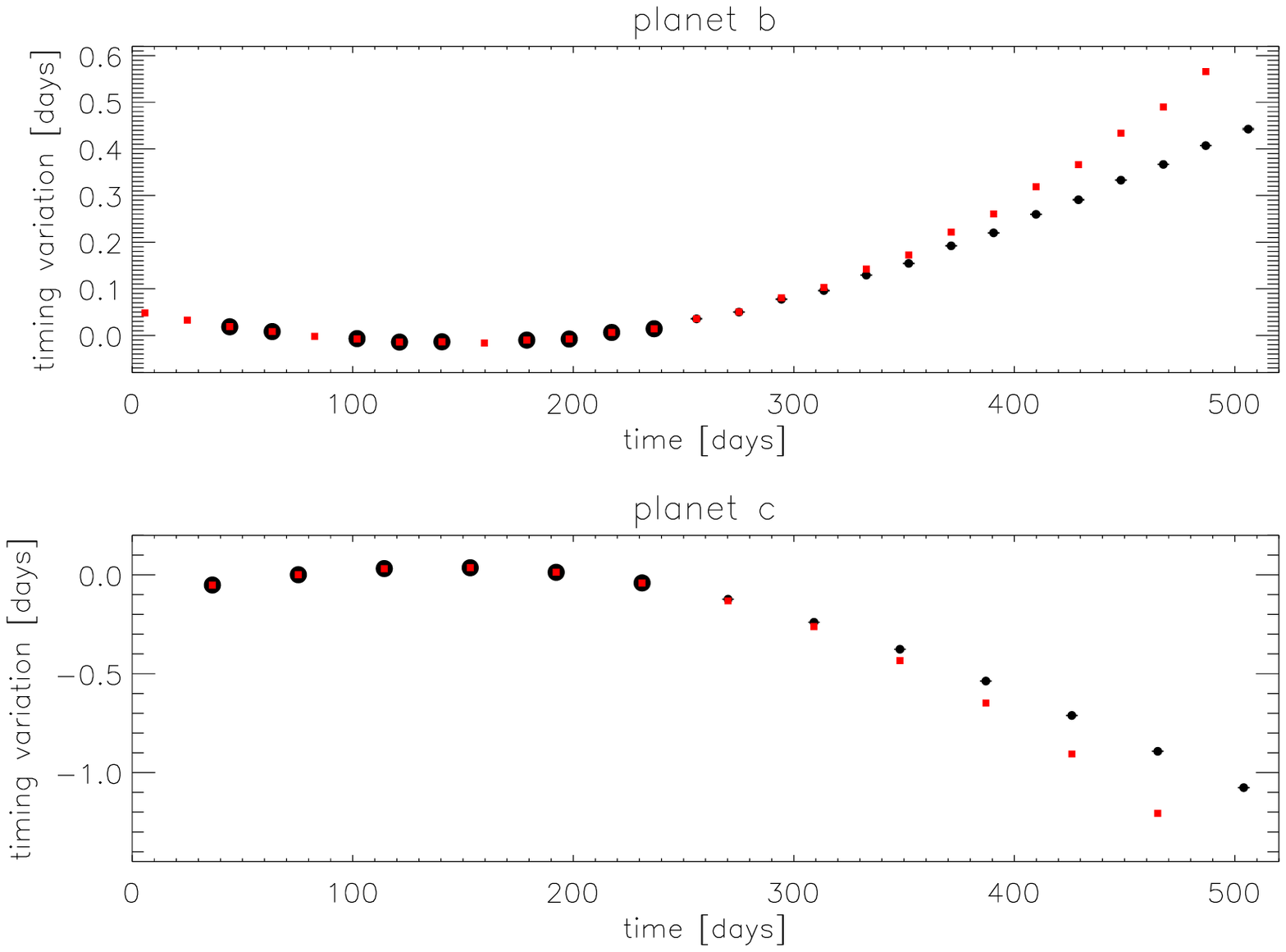}
\caption{Transit timing variation measurement, i.e. the difference
  between the observed transit times and a linear ephemeris, of planet
  b (top) and planet c (bottom). Data by H10 (large circles) compared
  to our best fit (red squares) obtained using only these early TTV
  measurements. Error bars are smaller than the symbols. The small circles 
  indicate the following transit times variations against the same linear 
  ephemeris as later obsered by Kepler. The time is given as
  BJD - 2454933.0.} \label{FigTTVOld} 
\end{figure}

\begin{figure}
\includegraphics[width=0.5\textwidth]{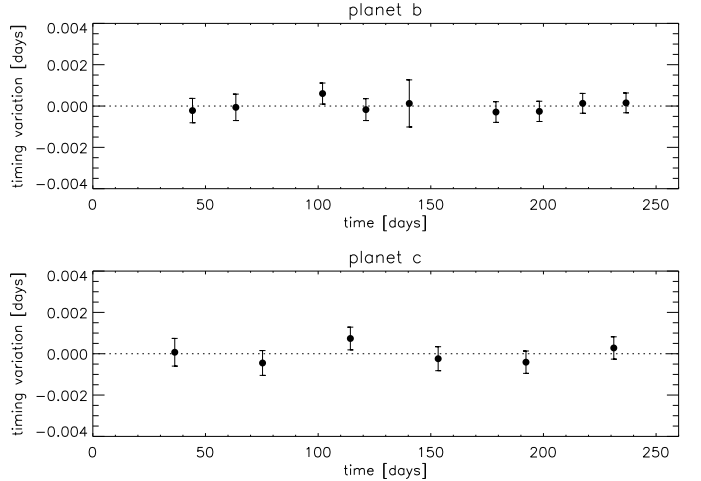}
\caption{Deviation between the measured transit timing variations of
  planet b (top) and planet c (bottom) planet by H10 and our best
  fit. The time is given as BJD - 2454933.0.} 
\label{FigTTVresOld}
\end{figure}

Our best fit parameters for the partial data set are summarized in
Table\,\ref{ResultsOld} and compared to pervious results. Our error
bars are taken from the distribution of parameters as shown in
Figures\,\ref{FigMassP1Old} and \ref{FigMassP2Old}. The best fit
model compared to the RV data is shown in Figs.\,\ref{FigRV}. The
comparison with Fig.\,(3) of H10 shows an nearly identical situation:
The observed RV variations can be matched reasonably well, however,
the deviation between the model and observation is up to 8\,m/s and
larger then expected given the error bars. Like in H10, we also
conclude that more RV observations would be necessary to check for the
influence of additional planets or stellar activity jitter. In
Fig.\,\ref{FigTTVOld} we show the TTV data together with our best fit
of the partial data set and in Fig.\,\ref{FigTTVresOld} we show the
residuals to that fit. The reduced $\chi^2$ of 2.4 is dominated by the
deviations from the RV measurements while the TTV measurements can be
matched within their error bars.

Basically, we can recover the literature values from H10 for the planetary
masses and the mass ration which agree well within the error
range. Also, the eccentricities and arguments of periastron agree
within the error range. In the other orbital
parameters, we find some discrepancies. We find a significantly lower
orbital period for planet b but a higher for planet c.
 
The fit to the partial data is also extended towards later
measurements (Fig.\,\ref{FigTTVOld}) and compared with the actual
transits observed by Kepler. About 250 days after the last transit
reported by H10 the deviation is already 0.1 days for planet b and 0.3
days for planet c. Given the small error bars, it is clear that the
planetary parameters have to be re-determined using the full Kepler
data set.

\section{TTV analysis using the full data set}
\label{FullTTV}
\begin{figure}
\includegraphics[width=0.5\textwidth]{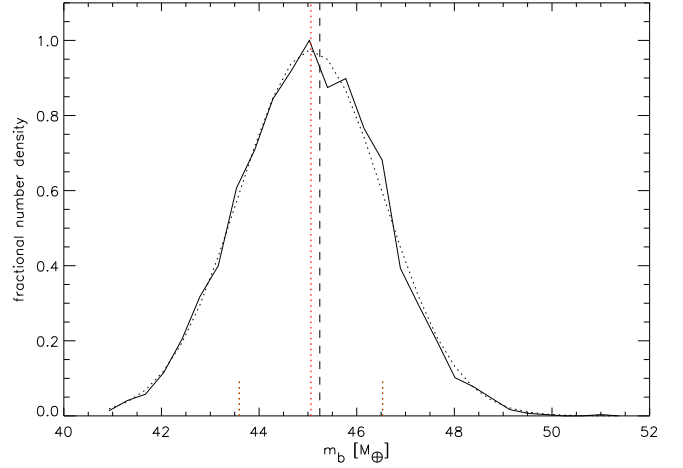}
\caption{Distribution of derived mass of planet b using the full data
  set (black solid line). The
   Gauss fit to this distribution is shown as black dotted line. The
   best fit as well as the $1\sigma$ error range is indicated as red
   dotted line. The median of the distribution is indicated as black
   dashed line. }
\label{FigMassP1}
\end{figure}

\begin{figure}
\includegraphics[width=0.5\textwidth]{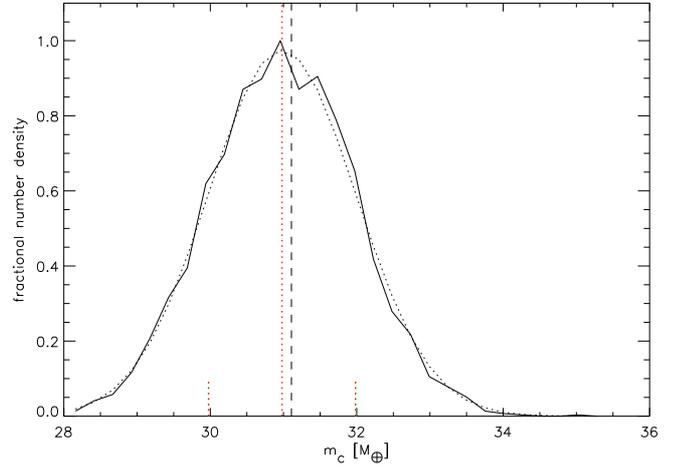}
\caption{Distribution of derived mass of planet c using the full data
  set  (black solid line). See also Fig.\ref{FigMassP1}.}
\label{FigMassP2}
\end{figure}

We repeated the analysis using the full data set in the same way as
described in the previous section, except that we now allow for non
coplanar orbits and increased the number of random starting values to
3000 (see Figures\,\ref{FigMassP1}, and
\ref{FigMassP2}). The full data set also allows to constrain the
planetary masses without including the RV data in the fit. As expected from the poor agreement between the
extrapolated solution from the partial data set, the planetary
parameters changed. The main change is a significant reduction of the
planetary masses in our new fit, while the mass ratio agrees within
the error range. In Table\,\ref{ResultsNew} we compare
the new parameters to those we obtained from the partial data set
(Table\,\ref{ResultsOld}). The increase in the orbital period for
planet b and the decrease for planet c are expected from the
deviations seen in Figure\,\ref{FigTTVOld}. 

Combining the fitted parameter $a/R_\star$ from
Table\,\ref{PhotometryTable} with the orbital semi major axis from the
TTV fit (Table\,\ref{ResultsNew}), we derive the stellar radius. Using
this in combination with $r/R_\star$ (Table\,\ref{PhotometryTable})
provides the planetary radius, which then can be combined with the
planetary masses (Table\,\ref{ResultsNew}) to the mean planetary
densities (Table\,\ref{ResultsNew}, lower block). 

\begin{figure}
\includegraphics[width=0.5\textwidth]{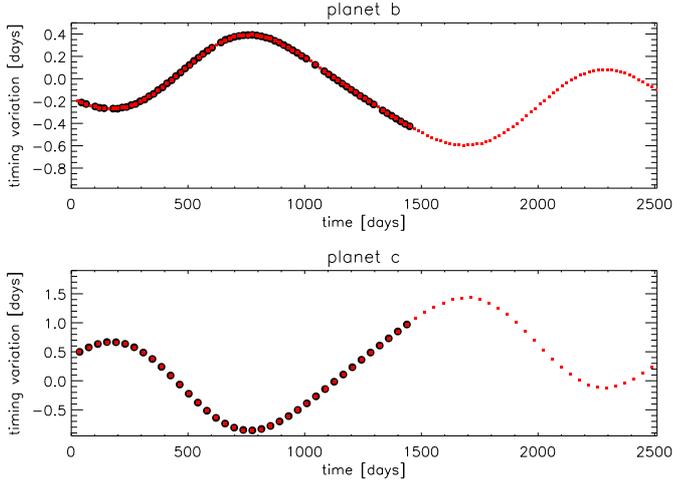}
\caption{Transit timing variation measurement, i.e. the difference
  between the observed transit times and a linear ephemeris, of
  planet b (top) and planet c (bottom) planet (large circles) compared
  to our best fit (red squares) obtained using the full Kepler data
  set. The error bars are smaller than the size of the symbols. The small 
  circles indicate the following transit times
  variations against the same linear ephemeris. The time is given as
  BJD - 2454933.0.} 
\label{FigTTV}
\end{figure}

\begin{figure}
\includegraphics[width=0.5\textwidth]{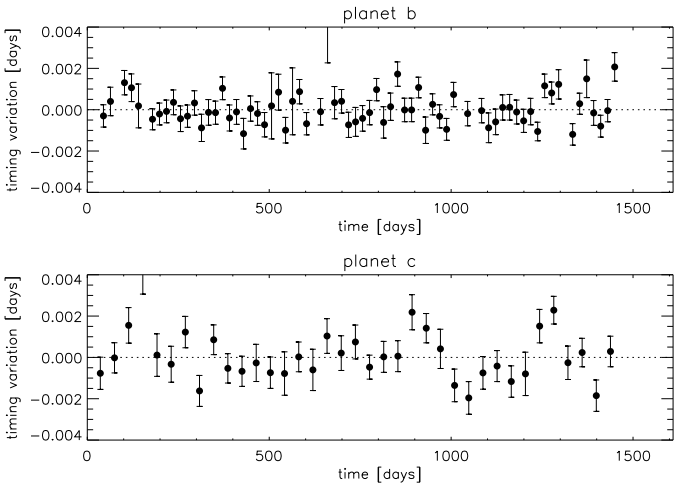}
\caption{Deviation between the measured transit timing variations of
  planet b (top) and planet c (bottom) planet and our best fit using 
  the full data set. The
  time is given as BJD - 2454933.0.} 
\label{FigTTVres}
\end{figure}

\begin{figure}
\includegraphics[width=0.5\textwidth]{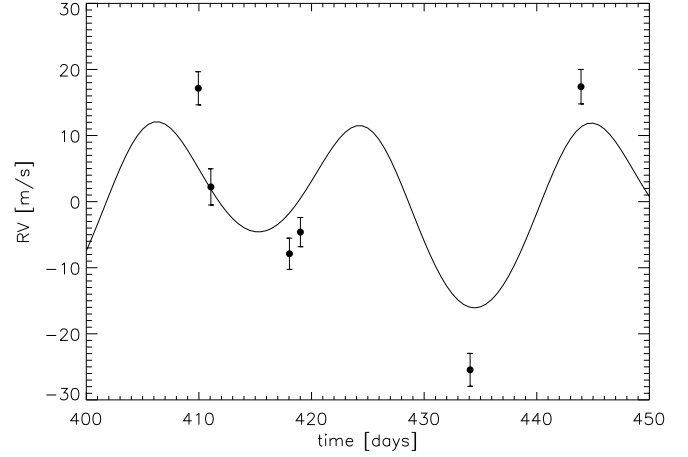}
\caption{Radial velocity measurements from H10 compared to those predicted 
  by our best fit using the full data set. We note that we do not include
  the RV data into our fit procedure but just derive it from the best fit 
  model. The time is given as BJD - 2454933.0.} 
\label{FigRVNew}
\end{figure}

\begin{figure}
\includegraphics[width=0.5\textwidth]{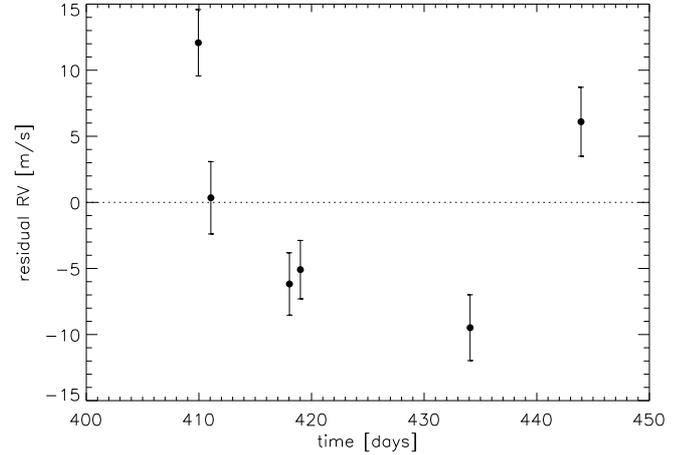}
\caption{Residuals of radial velocity measurements from H10 compared
  to those predicted by our best fit using the full data set. The time
  is given as BJD - 2454933.0.}
\label{FigRVNewres}
\end{figure}

Comparing the observed and calculated transit times in
Fig.\,\ref{FigTTV} now shows a good match over the whole observing
period. This can also be seen from the 
residuals (Fig.\,\ref{FigTTVres}) as well as from the reduced $\chi^2$
of 1.81. The lower planetary masses, however, lead to a less good match
of the RV variations (Figures \ref{FigRVNew} and
\ref{FigRVNewres}). We now have residuals of up to 12\,m/s. Note,
however, that the H10 masses were determined using the RVs in the fit, 
minimizing the RVs residuals by construction, whereas our full-dataset model 
did not fit for the RVs data. 

\begin{table}
\caption{Fitted parameters (upper block) of the planetary system of Kepler-9 derived from
  the TTV analysis using the full Kepler data set. The osculating
  orbital elements are given at a reference time BJD=2454933.0. The stellar mass
  is an input parameter with a distribution according
  \citet{2011A&A...531A...3H}, i.e. our errors take the uncertainties in the stellar mass into account. For
  comparison, we list our best fit parameters based on the TTV
  measurements from H10 (see Table\,\ref{ResultsOld}). Note that those
  orbital elements are given at a reference time BJD=2454900.0. We also list
  derived quantities (lower block) for the stellar radius, the
  planetary masses, radii, orbital periods, and densities using the fitted parameters $a/R_\star$ and
  $r/R_\star$ from Table\,\ref{PhotometryTable}.}
\label{ResultsNew}
\begin{tabular}{l l l l l}
\hline
\hline
Parameter & \multicolumn{2}{c}{this work}     & \multicolumn{2}{c}{this work} \\
          & \multicolumn{2}{c}{full data set} & \multicolumn{2}{c}{partial data set} \\
          & \multicolumn{2}{c}{without RV} & \multicolumn{2}{c}{with RV} \\
          & best fit      & $\sigma$          &  best fit & $\sigma$ \\    
\hline
m$_\star$ [M$_\odot$]       & 1.05 &  0.03 & 1.05     &   0.03 \\
\hline
m$_{\rm b}$/m$_\star$ [M$_\oplus$/M$_\odot$]
                         & 43.0       &  0.7     & 75.4& 3.3 \\
m$_{\rm c}$/m$_{\rm b}$     & 0.6875     &  0.0003  & 0.69& 0.004\\
P$_{\rm b}$ [days]         & 19.22418   & 0.00007 &  19.2159 &   0.0008 \\
P$_{\rm c}$ [days]         & 39.03106   & 0.0002  &  39.084  &   0.003 \\
e$_{\rm b}$                & 0.0626     & 0.001   & 0.10     &   0.02 \\
e$_{\rm c}$                & 0.0684     & 0.0002  & 0.08     &   0.02 \\
$i_{\rm b}$ [$^\circ$]      &  87.1      & 0.7     & not fitted &  \\
$i_{\rm c}$ [$^\circ$]      &  87.2      & 0.7     & not fitted &  \\
$\omega_{\rm b}$ [$^\circ$] & 356.9      & 0.5     & 357.5     &  21.0 \\
$\omega_{\rm c}$ [$^\circ$] & 169.3      & 0.2     & 101.5     &   4.0 \\
M$_{\rm b}$ [$^\circ$]      & 337.4      & 0.6     & 105.3     &  23.1 \\
M$_{\rm c}$ [$^\circ$]      & 313.5      & 0.1     &  36.6     &  20.6 \\
\hline
\hline
r$_\star$ [R$_\odot$]       & 1.23 &  0.01 & & \\
\hline
m$_{\rm b}$ [M$_\oplus$]    & 45.1 &  1.5  &  79.6     &   3.6  \\
m$_{\rm c}$ [M$_\oplus$]    & 31.0 &  1.0  &  54.8     &   2.6  \\
r$_{b}$ [R$_\oplus$]       & 11.1 &  0.1 & & \\
r$_{c}$[R$_\oplus$]        & 10.7 &  0.1 & & \\
r$_{d}$[R$_\oplus$]        & 2.0 &  0.02 & & \\
a$_{\rm b}$ [AU]          & 0.143 &  0.001& 0.143     &   0.001 \\
a$_{\rm c}$ [AU]           & 0.229 &  0.002& 0.229     &   0.002 \\
a$_{\rm d}$ [AU]           & 0.0271&  0.0001&           &       \\
$\rho_{b}$ [g\,cm$^{-3}$]  & 0.18 &  0.01 & & \\
$\rho_{c}$[g\,cm$^{-3}$]   & 0.14 &  0.01 & & \\
\hline
\end{tabular}
\end{table}


\begin{table}
\caption{Predicted future transit time for planet b and planet c from our
  best fit using the full data set. The time is given as
  BJD - 2454933.0}
\label{TTVFuture}
\begin{tabular}{l l l l }
\hline
\hline
\multicolumn{4}{c}{planet b} \\        
\hline
  1468.30749 &   1853.21471 &   2238.70993 &   2623.31170 \\        
  1487.53311 &   1872.48347 &   2257.96478 &   2642.53063 \\        
  1506.76023 &   1891.76418 &   2277.21419 &   2661.74940 \\        
  1525.98749 &   1911.03707 &   2296.46244 &   2680.96834 \\        
  1545.21757 &   1930.32080 &   2315.70445 &   2700.18775 \\        
  1564.44709 &   1949.59670 &   2334.94627 &   2719.40692 \\        
  1583.68102 &   1968.88203 &   2354.18160 &   2738.62704 \\        
  1602.91356 &   1988.15970 &   2373.41752 &   2757.84656 \\        
  1622.15229 &   2007.44513 &   2392.64722 &   2777.06742 \\        
  1641.38867 &   2026.72326 &   2411.87806 &   2796.28738 \\        
  1660.63310 &   2046.00730 &   2431.10335 &   2815.50893 \\        
  1679.87414 &   2065.28456 &   2450.33008 &   2834.72942 \\        
  1699.12498 &   2084.56572 &   2469.55214 &   2853.95164 \\        
  1718.37136 &   2103.84075 &   2488.77575 &   2873.17278 \\        
  1737.62904 &   2123.11760 &   2507.99568 &   2892.39566 \\        
  1756.88121 &   2142.38911 &   2527.21710 &   2911.61766 \\        
  1776.14570 &   2161.66033 &   2546.43584 &   2930.84130 \\        
  1795.40378 &   2180.92716 &   2565.65585 &   2950.06446 \\        
  1814.67463 &   2200.19167 &   2584.87413 &   2969.28904 \\        
  1833.93835 &   2219.45285 &   2604.09336 &   2988.51383 \\        
\hline
\multicolumn{4}{c}{planet c} \\        
\hline
  1477.96294 &   1867.51356 &   2255.75156 &   2646.06579 \\        
  1517.01523 &   1906.32679 &   2294.69070 &   2685.14079 \\        
  1556.05565 &   1945.12525 &   2333.66139 &   2724.21523 \\        
  1595.07993 &   1983.91531 &   2372.66051 &   2763.28891 \\        
  1634.08355 &   2022.70358 &   2411.68358 &   2802.36177 \\        
  1673.06218 &   2061.49673 &   2450.72550 &   2841.43373 \\        
  1712.01228 &   2100.30141 &   2489.78121 &   2880.50445 \\        
  1750.93174 &   2139.12401 &   2528.84626 &   2919.57315 \\        
  1789.82030 &   2177.97032 &   2567.91703 &   2958.63851 \\        
  1828.67971 &   2216.84512 &   2606.99083 &   2997.69845 \\        
\hline
\end{tabular}
\end{table}

The discrepancy of the observed and calculated RV variations as well
as possible slight systematic residuals in the TTVs of planet c raised
the question whether or not we can find evidence for a forth, possibly
non-transiting planet in the system. We therefore first searched for
additional transiting planets in the system in the photometric model's
residuals using Optimal BLS \citep{2014A&A...561A.138O} but found
none. We also repeated the dynamical analysis adding an outer,
co-planer, plane to the system. Since the parameter range for an outer
planet is huge, we restricted our search to orbits of the test planet
in 3:2, 2:1, 5:2, and 3:1 mean motion resonances to planet c. No
solution with a better reduced $\chi^2$ could be found. We note that
the short time span of the RV data -- less than one orbit of Kepler-9c
-- severly limit the orbits that one can hope to fit to such a test
outer planet.

Additionally, we have also checked our assumption that planet d has no
impact on our results: We included planet d in the dynamical model by
fixing its orbital period at the measured value, assuming a co-planar
and circular orbit, the latter motivated by the short circularization
time scale at the small orbit distance, and determined the mean
anomaly at the beginning of the integration to match the measured
transit time $T_{mid,3}$. We then refitted the full TTV data set for
planet d in the mass range of 1 to 30 Earth masses letting the
parameters of planet b and c re-adjust. We find a very shallow
$\chi^2$-minimum around 10,M$_\oplus$ but with insignificant
improvement compared to the tested mass range for planet d. The
parameters of planet b and c are unaffected within their error
bars. We conclude that we cannot derive any meaningful constraints on
the mass of planet d and find our solution sumarized in
Table\,\ref{ResultsNew} unaffected.

We conclude that neither an additonal outer planet in a low-order
resonant orbit nor the inclusion of planet d can improve on the very
systematic residuals of the RV signal. A more extended RV follow-up
would be necessary in order to come to a more conclusive result.

\section{Discussion and conclusion}
\label{Discussion}
\subsection{Partial vs. full dataset}

We re-analysed the Kepler-9 system using both the partial Kepler data
set that was available to H10 and the full data set available
today. The comparison between the previous and new results show, that
a very good fit to a planetary system in first order mean motion
resonance can be misleading if only a fraction of the interaction
time scale is covered. Even the much longer currently available Kepler
data set might not be sufficiently long for that. We therefore follow
H10 and extrapolated our best fit model into the future
(Fig.\,\ref{FigTTV} and Table\,\ref{TTVFuture}). Given the large TTVs,
ground based observations even with a marginal detection of the
transit should be able to check the solution proposed in this work.

H10 could confirm the Kepler-9b and Kepler-9c as planets from
photometry alone, but could only place weak constraints on their
masses without using RV data. They therefore included a few RV
measurements in their fit, and it comes as no surprise that the RV fit
is good since the partial photometry of the time did not have the
constraining power to match the RV data. They also predicted that
future Kepler data would be more constraining of the planetary masses,
and indeed our results have smaller formal error bars on both planets'
masses from photometry alone. We note, however,that the systematic
residuals shown in Fig.\,\ref{FigTTVres}, and especially
Fig.\,\ref{FigRVNewres} cause us to warn of unmodeled phenomena, such
as other planets in the system or longer time-scale interaction between
the planets or stellar activity.

\subsection{The revised planets}
The scaled radii r$_{b,c}/R_*$ we determined are slightly larger than the ones
obtained by H10 by $\sim 3 \sigma$ and $\sim 4.5 \sigma$ for Kepler-9b
and Kepler-9c, respectively. The new values are much more constrained
with formal errors 5 to 8 times smaller. Actually, Kepler's data
allows in principle to determine the planets' mass to $2.8\%$ and the
planets' radii to better than 0.2\% -- but those are limited by our
knowledge of the host star properties. Furthermore, Kepler-9 was
measured in short cadence mode (1 minute sampling instead of the
regular 30-minute sampling) starting from Quarter 7, which allows for
an even better timing precision (and thus mass determination). While we
did not use short cadence data, using this data would have had little
effect on the global uncertainty which is dominated by stellar
parameters errors.

The newly determined masses and radii of Kepler-9b and -9c change the
nature of these planets relative to the one described in H10. Both
planets are now detetmined to have size similar to Jupiter's but they
are 7 to 10 times less massive than Jupiter, i.e. have densities about
1/3 of the density given in H10. Consequently, both planets have very
low derived densities of $\rho_b\sim=0.18\,g\,cm^{-3}$ and
$\rho_c\sim=0.14\,g\,cm^{-3}$ -- among the lowest known. H10
specifically excluded coreless models for the planets, but the more
abundant data we have today forces us to reconsider that Kepler-9b and
-9c may not have cores at all. This result is of special interest in
the context of the core accretion theory \citep{1996Icar..124...62P}:
with masses of 30.6 and 44.5 $M_\oplus$ these planets have apparently
just started their runaway growth when it stopped at this relatively
rare intermediate mass.

\begin{figure}
\includegraphics[width=0.5\textwidth]{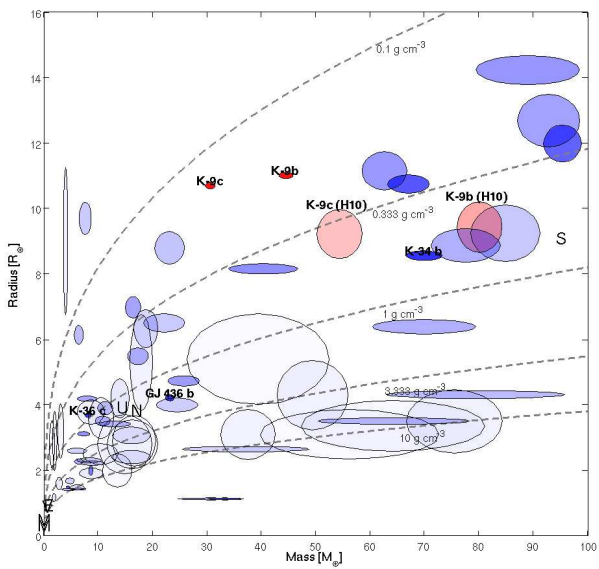}
\caption{
The mass-radius distribution of all well determined planets (both mass
and radii determined to better than 3$\sigma$). For each planet the
mass- and radius- semi-major axes represent the 1-$\sigma$ error bar,
and the transparency is such that better determined planets are more
opaque. Contours of constant bulk density are shown in dashed gray
lines. The names of some of the better-determined planets are
indicated. All planets are shown in shades of blue, but Kepler-9 which
is shown in shades of red: larger (and more transparent) symbols for
the H10 values, and smaller (and more opaque) symbols for the current
study's values. Solar system planets are shown as letters.}
\label{FigRM.ps}
\end{figure}

Figure\,\ref{FigRM.ps} shows the masses and radii of lower-mass
($M<100 M_\oplus$) planets that have both mass and radius known to
better than $3\,\sigma$ \footnote{Extracted from the NASA Exoplanet
Archive (\texttt{http://exoplanetarchive.ipac.caltech.edu/}) on
January 21, 2014}. It is evident that the new locations of Kepler-9b
and -9c put them at the edge of the mass-radius distribution, with
very low density and in a mass range that is very poorly sampled, and
yet -- both planets are now among the best-characterized exoplanets
known with bulk densities known to $7\%$ or better. The recent
successful launch of the GAIA mission further highlights that last
point: the knowledge about both Kepler-9b and -9c in both radius and
mass is limited by the knowledge about their host star. GAIA's
observations will fix Kepler-9's properties to high precision,
allowing to use other data (such as the available short cadence data)
to further reduce the uncertainty on the physical parameters of
Kepler-9b and -9c, and significantly so.

Finally, we note that Kepler-9d is now determined to have a radius of
$2.00 \pm 0.05 R_\oplus$, an increase relative to H10. The increased
size, together with the low metal content of its neighboring planets,
suggest that Kepler-9d may not be rocky, or at least that it may 
have a significant volatiles fraction, again unlike the initial
suggestion by H10. If this is true, then Kepler-9d is perhaps similar
to the new and exciting subgroup of low-mass low-density planets
\citep[e.g. Kepler-87c or GJ\,1214][]{2014A&A...561A.103O,2009Natur.462..891C,2013ApJ...775...80F}

\section{Acknowledgments}
A.O. acknowledges financial support from the Deutsche
Forschungsgemeinschaft under DFG GRK 1351/2.





\end{document}